Modelling transport provision in a polycentric mega city region.

Florent Le Néchet

Université Paris-Est, Laboratoire Ville Mobilité Transport, UPEMLV : 5 boulevard Copernic, Cité Descartes F 77454 Marne-la-Vallée cedex 2 France

florent.lenechet@u-pem.fr

## Abstract

The aim of this paper is to present a model of interaction between transport and land use which aims at endogenously integrates the provision of transportation infrastructure and its effects on land use, with a long term perspective (Lowry, 1964, Wegener, 2004, Levinson, 2011, Bretagnolle, 2014, Mimeur *et al.*, 2015). LUTECIA (Land Use, Transport, Evaluation of Cooperation, Infrastructure provision and Agglomeration effects) model puts emphasis on multiscale processes of urban growth, in the context of the emergence of Mega-City Regions (MCR, Hall & Pain, 2006). It allows us, in this exploratory phase, to characterize via simulation the conditions of development of polycentric metropolises. Nevertheless, we argue that such approaches are all the more necessary that we are in a period of multiple transitions that classical modelling tools have difficulty to capture.

## Introduction

The aim of this paper is to present a model of interaction between transport and land use which aims at endogenously integrates the provision of transportation infrastructure and its effects on land use, with a long term perspective (Lowry, 1964, Wegener, 2004, Levinson, 2011, Bretagnolle, 2014, Mimeur *et al.*, 2015). LUTECIA (Land Use, Transport, Evaluation of Cooperation, Infrastructure provision and Agglomeration effects) model puts emphasis on multiscale processes of urban growth, in the context of the emergence of Mega-City Regions (MCR, Hall & Pain, 2006). It allows us, in this exploratory phase, to characterize via simulation the conditions of development of polycentric metropolises.

Cities are said to be complex (Benenson & Torrens, 2004; Batty et al., 2012), in the sense that the modification of a component of one of the subsystems would potentially induce changes to all the components of all the subsystems. LUTECIA model focus on the coevolution of location system and transportation system (Raimbault, 2018).As an illustration of such interactions, let us mention the rebound effect of travel speed on trip distances (Litman, 2017a). At metropolitan level, gains in accessibility due to highway construction and thus improved speeds overall resulted in unforeseen urban sprawl (Banister, 2011). Complex imbrication also occurs between multiple spatial scales of planning: decisions of transport provision made at a local scale can have repercussions at a wider scale, and vice versa. For instance, Appert (2004) showed how the Green Belt policy in London (belt of restricted urbanization located circa 30 km from the city center), a measure which was dedicated to limit urban sprawl, created in fact leapfrog sprawl and increased travel distance for new residents located outside the green belt and commuting to London. Conversely, the spatial organization at a metropolitan level can influence the need for transport provision: the proximity of two cities and increase of commuting flows can over time create needs for the development of efficient transportation links linking the two cities (Le Néchet, 2012)

This article focuses on the complex coevolution between decision making of transport infrastructure provision operated by local and metropolitan stakeholders and urban dynamics hence changes in the spatial organization of population and activities densities. Land use and transport infrastructure networks come from planning decisions made over a long period of time by a diversity of actors at individual and collective levels (Venables, 2007; Vickerman, 2017). Individuals make location choices based on several criteria (for instance, housing affordability, accessibility to workplaces). At the collective level, urban actors play a role in the development of cities, such as national or supra-national bodies and private actors, capable of making heavy investments.

The contribution of this article lies in the exploration of paths of urban dynamics with evolving transportation network, and of the conditions of the emergence of governance at MCR level. We start be a short discussion on the emergence of Mega-City Regions (Hall & Pain, 2006), then raising the question of the level of governance most fitted to transportation issues. In the last section, we introduce the theoretical agent-based model used to explore such questions.

## Emergence of Mega-City Regions

This section describes the emergence of MCR, in Europe and the United States. Many examples are studies in Europe (PolyNet research network for instance), including the well-known Randstad Holland (Lambregt & Kloosterman, 2012). During the second part of the twentieth century, the reduction in transport costs has been accompanied by a double dynamic of urban sprawl (Le Néchet, 2015) and hierarchization of the system of cities (Bretagnolle et al., 2007; Pumain et al., 2015). Intra-urban and inter-urban dynamics tend to overlap: (Champion, 2001; Pumain et al., 2006). As stat-ed by Soja (2011) « The urban, the metropolitan, and the subnational-regional scales seem to be blending together ». Mega City Regions (Hall & Pain, 2006) are geographical objects lying somewhere between cities and system of cities, structured by flows and networks of communication.

Territories of several thousands of square kilometers of discontinuous urbanization are now functionally integrated, with various kind of flows (business & leisure trips, Scott, 1997, logistics supply chains, Heitz & Dablanc, 2015, long-distance commuting trips, Conti, 2015). Several geographical levels are now *de facto* interrelated in the fabric of cities (what Soja, 2011 call "multilevel urbanization processes"). In particular, these evolutions the questions of which type of governance are more adequate to such geographical objects to achieve social, economic or environmental goals. The links between the reconfiguration of flows within territories and reconfiguration of governance at the metropolitan or regional level are though at different scales, and constitute an open and difficult question (Lefèvre, 2009; Cowell, 2010; Le Néchet, 2017). The literature does not offer a consensus over the evaluation of the resulting synergies (Meijers, 2004, Davoudi, 2007) of such metropolitan integration policies, but the question of the main scale of decision is little raised by the literature. As an example, let us compare Paris (France) and Rhine- Two research questions are treated at this stage: what is the impact of the spatial organization of densities (polycentric versus monocentric) on the emergence of Mega-City Regions? What is the influence of governance system (centralized versus noncentralized) on the level of polycentricity of the Mega-City Region?

## *Urban sprawl, functional integration and governance levels*

Ruhr area (Germany), two European metropolitan areas of relatively similar total population (12M) and size (12,000 km²), with very contrasted spatial organization. Figure 1 illustrates how different these networks are, that have emerged over the long term.
In Paris, the local network ("métro") is of small spatial scope, which arises from the scale of decision before the construction of the network : in 1900, the municipality has been able to impose its project to the national level (Larroque et al., 2002). Thus, two scales of network emerged, still visible today, which is not the case for example in another monocentric city like London. As the cities of the Rhine-Ruhr region (Köln, Düsseldorf, Essen, Dortmund, Duisburg) have developed in relative independence over time, before a gradual and incomplete functional integration (Blotevogel, 2001, Le Néchet, 2012), it is logical to observe here also two scales in the transportation network, the regional network "S- Bahn" having been organized since 1967.

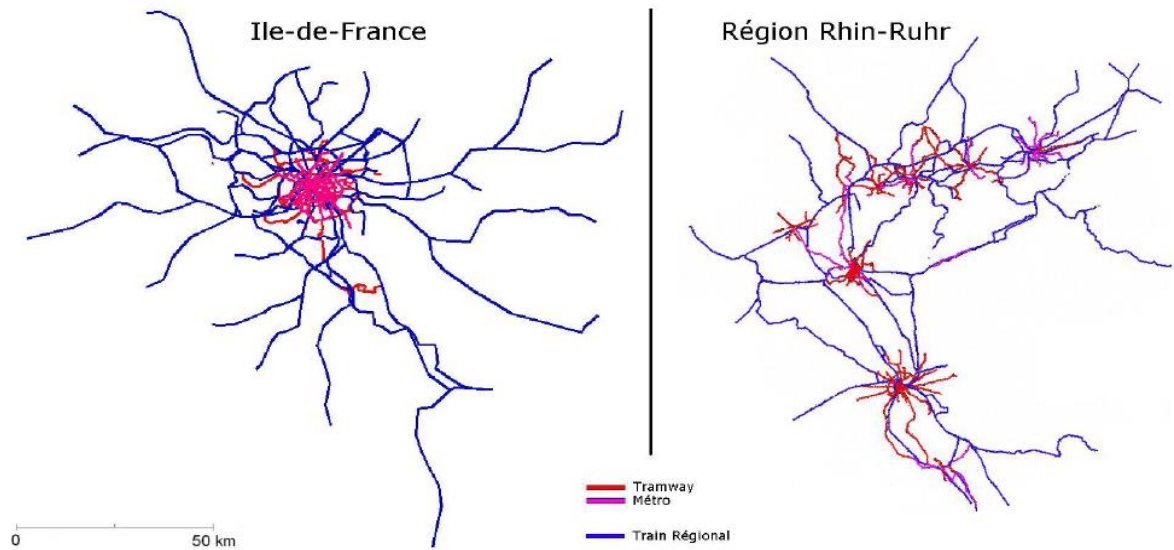

**Figure 1: Transit networks, at several spatial scales, in Paris Metropolitan Area "Ile-de-France-Paris", France (left) and in the Rhine-Ruhr region, Germany (right).**

## *Research question : multilevel governance and coevolution transport and land use*

In this article, we start from this very simple question: why are transport networks organized the way they are? Since the time scales involved are long (railway networks back from the beginning of the 19th century, the motorway networks of the early 20th century), this question invites us to think more generally about the interactions between transportation networks on one hand, and demographic evolution on the other hand. Kasraian et al. (2016) provide an extensive meta-analysis which highlights the complexity of these links. Most of the studies in this field use the notion of accessibility (Hansen, 1959) to understand and predict the effects of a transport infrastructure on territorial development, and in a more systemic approach, as an operative concept for studying the coevolution dynamics between transport and territories (Raimbault, 2018). To add to this complexity, transport infrastructures are decided and financed by a multiplicity of territorial stakeholders, since they potentially affect multiple territorial scales, from local to national or even international scale for fast networks such as high-speed railways or airports. Due to the polycentric nature of the MCR, our research question is thus transformed into the following one : what is the impact of a type of governance, hence giving the advantage to the local or to the whole region, on the dynamics of coevolution between transport and spatial distribution of the demography? This question takes place in the current debates about the formation of governance bodies at metropolitan level (Heeg et al., 2003 ; Douay, 2010) or at MCR level (Le Néchet, 2017). In this article, we tackle these questions via simulation.

According to our hypotheses, local stakeholders will tend to favor the accessibility of workers living within their governance area. Conversely, a regional stakeholder would aim at better global accessibility of the whole territory, which would often lead to different choices. Overall, one would expect MCR governance integration to occur when regional planning significantly outpaces the sum of local planning outcomes in terms of total accessibility (even if other factors might affect integration of governance, such as cultural differences, Blotevogel, 2001). Within the framework of simulation, to account for the rationality of territorial stakeholders at various spatial scales (cities, region), we are inspired by the paradigm Belief Desire Intention (Taillandier et al. 2011): actions of a territorial stakeholder will be driven by the perception of the elements of the system that he is likely to act on.

However, most models of literature do not raise the issue of multiple territorial stakeholders. In most cases, choices are made only at the level of economic agents (individuals, real estate developers, etc.), or they are made according to a global rationality, implicitly assuming that one single actor has ability to take all decisions for a whole region. The reality is more complex: individual rationalities and the concept of global rationality come up against the rationalities of multiple territorial actors operating on multiple scales (Trigalo, 2000).

Adopting a modelling framework separating individual mobility and migration choices on one hand (micro scale) and stakeholder long-term investments such as transport provision on the other hand (macro scale), the following section focus on the literature review to understand agents choices at each scale.

## *Transport provision, accessibility, and residential location models*

Indeed, transport infrastructure provision follows various logics, for instance supporting land development (the Transit-Oriented Development Literature, Litman, 2017b) or improved accessibility for deprived neighborhoods (Desjardins & Drevelle, 2014). Acknowledging this complexity, a part of the literature focus on the possible effects of transport provision on to the spatial organization of densities (Banister, 2001; Litman, 2017b). LUTI (Land Use Transport Interaction) models extend classical four-stage model of transportation demand to take into account residential mobility. A key hypothesis of the model is that an increase in accessibility due to a new transportation link is likely to bring opportunities for new urban developments and/or an increase in residential or employment density around the new modes of transportation (Wegener, 2004). Most of such research are focused at the metropolitan level, only a few studies trying to assess the effects of a variation of boundaries on the residential choice model output (see Thomas, 2018 for a complete analysis on LUTI models).

Based on economic hypotheses such as monocentric model by Alonso (1964) and Lowry dynamic model (1964), some models link individual choices, whose determinants are synthesized by sophisticated mathematical functions via utility functions that result from these choices. Based on elements identified as exogenous (transport cost, available infrastructure, existing housing stock), and a finite number of selection criteria, it is possible to study the Pareto equilibrium configurations of the system (Delons et al., 2008). Some other have an agent-based approaches (Tannier et al., 2015). Scenarios can then be made to assess how a system would react in response to exogenous modifications such as an increase in the cost of mobility or a change in transport supply) This type of approach has proven useful in operational contexts to assess the socio-economic benefit of transport provision (Vickerman, 2017). However, data on the supply side are considered as exogenous: the model cannot predict the growth of infrastructure or other major structural changes (Raimbault, 2018). This constitutes the main limitation of this approach in the perspective of modeling the dynamics of cities in a longer term.

To understand the long-term evolution of urban systems, it is useful to have other tools. Literature on transport network evolution is scarce (Raimbault, 2018), and existing models often focus on the geometric aspects (for instance percolation models) and are rarely explicitely with the socio-demographic aspects of the territory. Articles by Zhang and Levinson (2007), Xie & Levinson (2009), or Cavailhès et al. (2010) are notable exceptions that endogenously account for the changing transportation networks in relation to the changes of land use and commuting flows. Our approach is comparable but extends to the case of multiple actors.

## Overview of the LUTECIA model

This paper introduces a LUTI model, developed and implemented on the Netlogo platform (Wilenski, 1999) and presented in Le Néchet (2010, 2017) and then Raimbault (2018) aimed at exploring the complex multiscale evolution of transportation networks and spatial structure, in the context of polycentric Mega City-Regions (MCR, Hall & Pain, 2006). The model combines a classical discrete choice LUTI model and a module that predicts endogenously the evolution of transportation network, given a configuration of territorial stakeholders respective powers.

In order to explore the formation of governance at MCR level over several decades. Two main hypotheses are used in the formalization of the model: firstly, the development of transport infrastructures affects the spatial organization of population and jobs; secondly, planning stakeholders are in interaction to decide for heavy transport provision investment.

A stylized city is proposed which consists of the following elements:
- The cells are the elementary zones which support housing and jobs
- The metropolis: set of cells constituting the closed environment in which population (only workers in this version) and workplaces take place.
- The workers are located on the cells and differentiated by socio-occupational categories.
- Jobs are also located on the cells; the types of jobs are associated with socio-professional categories (we used a simple typology)
- Planners are agents responsible for providing transport infrastructure. Two type of actors are implemented in the model: local stakeholders called "mayors", who divide the metropolis into M disjointed areas, and a metropolitan "governor". These (M + 1) agents have their own objective function when provisioning transport infrastructures.

The model is articulated around three sub-modules, ran successively during a time step:
- A transport module, which computes the demand for travel according to a given configuration of the internal distribution of workers and jobs, as well as the supply of transport infrastructure.
- A land use module that reallocate workers and jobs based on metropolitan accessibility via transportation network
- A governance module, which settles the decision making of transport infrastructure provision according to differentiated development rules at several scales.

To account for the hypothesis stated, the underlying principle of the model is as follows: a spatial configuration of transport infrastructures being given, workers and jobs relocate in space, according to the transport possibilities offered by the city. In return, depending on the location of workers and jobs, a new transportation infrastructure is built in order to improve the overall accessibility level. This growth of accessibility is non-linear: some infrastructures simply extend an existing line and reinforce an ongoing process of urban sprawl; conversely, when an infrastructure connects two major centers, a structural change occurs, with possible further integration towards a polycentric metropolis as described by Champion (2001).

In the initial configuration, workers and jobs, are spatially distributed exogenously. Workers are distributed according to Clark's (1951) empirical law: the monocentric model. The population density $\rho(c)$ in cell c, distant from from center $C_i$ by a distance $d_i$ is according to polycentric model of Heikkila et al. (1989): $\rho(c) = \sum_{i=1}^{M} A_i e^{-b_i d_i}$ with M centers.

This section details the equations used in the model, for each of the three modules presented. The model relies on a classical four-stage transport demand model, discrete choices model of location choices of households and jobs. The originality of the model is the inclusion of a model accounting for the construction of new transport infrastructures.

## *Transport submodel*

The four-stage model of transportation demand is commonly used in transportation engineering to model the demand for travel, once given location-based attributes (workers and jobs per municipality generating transport demand) and the attributes of transport infrastructure (travel time, capacity) summarizing the transport offer. In detail, the four stages conventionally implemented, with varying degrees of sophistication, are as described as suggested by Bonnel (2001):

1. Generation of the travel demand. The aim is to estimate, trips at origin and at destination, by travel zone, for various trip purposes. In this model, only commuting is taken into account, differentiated by socio-professional category.
2. Distribution of commuting flows, carried out according to the gravity model:

If $A_i$ and $E_j$ are respectively the number of workers of zone i, and the number of jobs of zone j, the number of displacements between the two zones $\Phi_{ij}$ is obtained by solving the system of coupled equations (Furness algorithm):

$$\forall i, j = 1..N, \Phi_{ij} = p_i q_j A_i E_j e^{-\lambda d_{ij}}$$

under constraints : $\forall j = 1..N, \sum_{k=1}^{N} \Phi_{kj} = E_j$ et $\forall i = 1..N, \sum_{l=1}^{N} \Phi_{il} = A_j$

given $\forall i = 1..N, p_i = \dfrac{1}{\sum_{l=1}^{N} q_l E_l e^{-\lambda d_{il}}}$ et $\forall j = 1..N, q_j = \dfrac{1}{\sum_{k=1}^{N} p_k A_{kl} e^{-\lambda d_{kj}}}$

$p_i$ and $q_j$ are temporary parameters; λ can be interpreted aversion to the distance of individuals; here $d_{ij}$ represents the travel time between the two zones, depending on the transport infrastructure present (if the "as the crow flies" – AFC - route induces a lower time, it is retained).

3. Modal choice. Here, only car use is implemented; therefore, this submodule is disabled.

4. Traffic assignment will compute the frequentation of the different sections of the transport networks. This step is performed by calculations of shortest paths, taking into account congestion, via an implementation of a static Dijkstra algorithm. Note that LUTECIA model differentiate "local" and "regional" roads: local roads are not represented in the model but correspond to AFC distance, they are no subject to congestion. Regional roads are the transportation links that are built throughout the model: they are subject to possible congestion (travel speeds depend on flow-capacity ratio according to the classical BPR 1964 function).

## *Location choice submodel*

The literature on residential mobility is mainly based on the idea that individuals and firms, in the choice of their location, promote greater accessibility to the desired resources as well as attributes of the place (Tannier et al., 2015). In LUTECIA model, no land auction is implemented but the local distribution of densities are accounted for in the relocation submodule, making it possible to implement the two opposite forces: density aversion and desire for accessibility.

This utility function depends on the position of the cell relative to its environment (Accessibility, noted $X_c$), and its own attributes (Urban Form, noted $F_c$).

The implementation adopted is as follows: a Cobb-Douglas function makes it possible to synthesize the utility function of the workers (and jobs) for each cell c, Uc.

$$U_c = X_c^{\gamma} F_c^{1-\gamma}$$

In a detailed way, the functions are calculated as follow (for the workers, the formalizations are symmetrical for the jobs):

$$X_c = \sum_s A_i^s \sum_j E_j^s e^{-\nu d_{ij}}$$

$$F_c = \prod_{s,s'} (A_i^s)(E_i^s)(A_i^{s'})^{m_{s,s'}} (E_i^{s'})^{m'_{s,s'}}$$

The parameter ν accounts for the cost of energy in the sense that greater distances can be more or less difficult to achieve on a daily basis depending on the city wealth and the technology of vehicles; this parameter is common to all agents of the model. Note that it is a strong assumption which can be relaxed in further versions of the model.

A proximity matrix between socioprofessional categories allows to account for the preference between workers and jobs in location choice model; three possibilities are implemented:
categories s and s' are indifferent to their mutual presence ($m_{s,s'} = 0$), avoid each other ($m_{s,s'} < 0$) or appreciate their mutual presence ($m_{s,s'} > 0$).

Once these utility functions have been determined, the workers and jobs are distributed on the grid, so that the best locations are the most chosen; a discrete choice model has been implemented: the probability for a worker (respectively a job) to move into cell c is as follows, where μ is the sensitivity of workers (respectively jobs) towards a differential of accessibility.

$$P(c) = \frac{e^{\mu U_c}}{\sum_{c'} e^{\mu U_{c'}}}$$

NB : il manque toute une partie sur l'évaluation socio-économique des infrastructures

## *Governance submodel*

Growth of a transportation network through coordination between several territorial stakeholders

Finally, a last step is implemented to allow for endogeneous modification of the transportation network. As stated earlier, we wish to include in the model the evaluations of transport provision needs from both local and regional stakeholder. Our modelling choices lie on three main hypothesis : (i) in general stakeholders will seek maximization of accessibility to jobs for its residents given a level of new transport resources ; (ii) due to historical pathways (Paasi, 1986), there is an external and somehow constant level of concentration of governance within the region, between two extreme cases : all decisions are taken at regional level, with no role to local stakeholders and all decision are taken at local city level, with no role for regional stakeholders ; a parameter can account for the proximity towards such extreme configurations ; (iii) : in case of local stakeholder having a say in transportation provision, the relative power of local stakeholder will be a function a their relative economic wealth.

Thus, due to hypothesis (iii) a local city has an implicit interest in developing urban amenities. In addition, the territories are in a situation of cooperation: the workers of each territory can use the infrastructures created, independently of the stakeholder that made the investment; this concurrence / cooperation processes appear realistic in a context where there are increasing interactions between spatial scales as mentioned by Parr (2004) and Cowell (2010).

The infrastructures are here developed one by one, by investments made 100% by one of the administrative authorities: local mayor or regional autority. This hypothesis indeed neglects part of the complexity of the political decision-making process, which often leads, in the case of transport infrastructure, to compromises carried by multiple actors (Ollivier-Trigalo, 2000), and to joint financing by different stakeholders. However, we argue that over the course of several iterations of the model, the various stakeholders will have had chance to develop infrastructure according to their utility function, proportionally to their respective importance.

Transport provision is implemented using a two-step approach: (i) The choice of stakeholders that will develop the infrastructure network: local decision (one of the mayors) or metropolitan decision? (ii)The choice of infrastructure: which criteria? which method is chosen?

(i) Firstly, randomly chosen decision will be at the local or metropolitan level, according to an exogeneous parameter, the share of local decisions ξ. Note that the determination of mayor is set exogenously, and do not vary during the simulation (there is no feedback loop from land use to the administrative division between stakeholders).
If the decision is local, another random step is achieved to determine which mayor will have the opportunity to decide on the construction of the new infrastructure. This random choice is conducted according to a probabilistic model taking into account the relative power of mayors through number of jobs in the mayor's territory (in order to take into account the relative wealth of each territory); In other words, there is a feedback loop of the distribution of employment on the propensity of each mayor to make a development decision. In detail, if $Y_i$ is the total number of jobs in Mayor i's territory, the probability that this mayor will make the decision is: $\varepsilon_i = \dfrac{Y_i}{\sum_{j=1}^{M} Y_j}$.

(ii) Once a territory T has been chosen on which the decision to build an infrastructure is based, the choice of infrastructure is made with a view to maximizing the accessibility of the T territory's workers to all metropolitan jobs. This choice is, of course, debatable, but it allows us to give a synthetic account of the desire usually expressed by planners and urban planners to encourage social exchanges between individuals and economic functioning of the territory. The current configuration is associated with a "current" accessibility $X(T) = \sum_{c \in T} X(c)$, where X (c) is the accessibility of the cell c, belonging to T.
If a new infrastructure, Z, is built, the accessibility function will, after application of the transport module, be modified: $X_Z$ (T). It's about retaining the infrastructure that maximizes new accessibility:

$Z^* = \arg\max_{Z} X^Z(T)$.. This research is carried out here for all the infrastructures connecting two neighboring zones, which implies substantial computing times, explaining why a small grid is used at this exploratory stage.

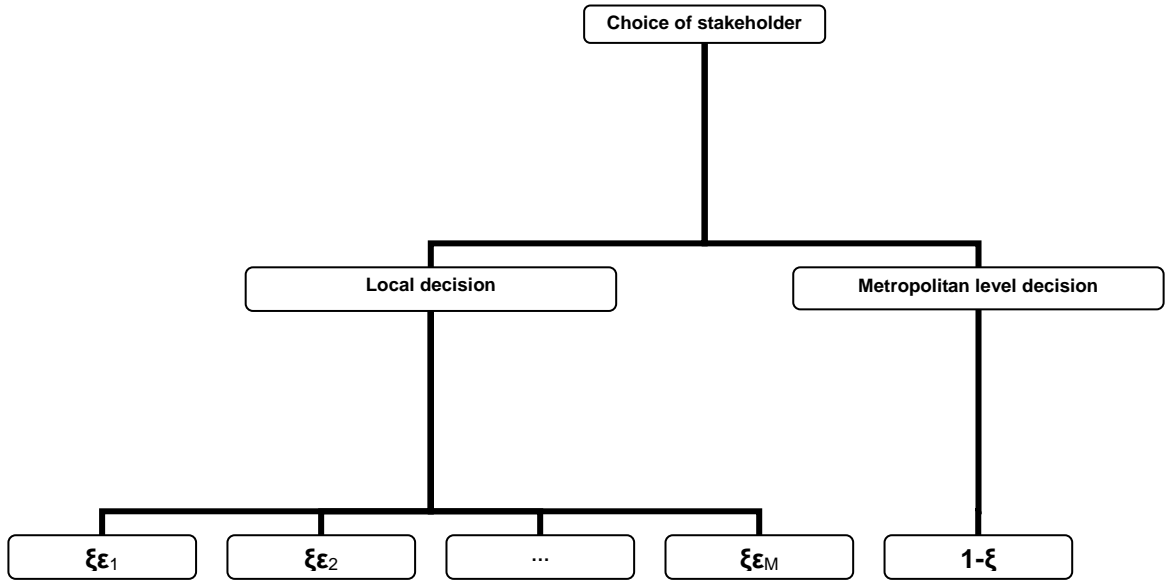

**Figure 2 : Probability for each territorial stakeholder to obtain the decision at each time step.**

As such, the metropolitan dynamic is the result of a stochastic process, which we explore in the next section.

## Results

In this last section, first applications of the model are proposed. Note that in this archive version of the paper, the validation roadmap is not detailed ; parameters used here derive from a "expert" validation in the sense that they were chosen to produce realistic urban dynamics (for instance, the speed of sprawl has to be controlled to ensure realistic response to provision of transport infrastructure). The following section describes the dynamics of the transportation network in absence of land use change, but with varying type of governance and different initial configurations in term of land use.

### *Growth of transportation network, no land use change*

Figure 3 shows the state of a simulation after a small number of steps; Mayor 0 (gray, bottom right)

and Mayor 1 (red, top left) develop the territory. The population density, in green, shows a city more populated than the other. It is a rather monocentric city, with a short distance between the two centers. Hence, Mayor 1 is more likely to apply its utility function to the transport infrastructure provision, compared to Mayor 0.

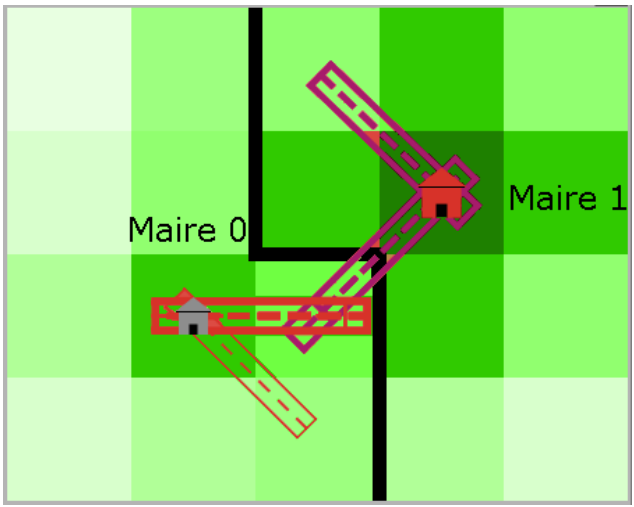

**Figure 3 : Simulation outputs after 3 timesteps : exclusively local decisions (left); exclusively metropolitan (right).**

Each simulation (construction of six infrastructures) is reproduced 30 times to account for the stochastic variability of the model. Figure 4 shows the mean values (and the variation ellipse) of the following indicators:
- Total accessibility,
- Total travel time of individuals

Given the spatial configuration, high accessibility level corresponds to a well-meshed network (facilitating exchanges between the two cities of the metropolis), and low travel times illustrate either a lower territorial integration, or an absence of congestion. The results indicate equivalent accessibility in the case where decisions are taken at the metropolitan level only or by the mayor of greatest importance, the mayor 1. However, when decisions are made by the other, the total time of travel is lower which indicates a lower degree of congestion. Conversely, when decisions are mainly made at Mayor level 0, total accessibility is lower. It is not a question of determining here an "optimal" governance system means, these two indicators having been retained among many others conceivable, but of exploring the articulation between scales in this model transport / land use

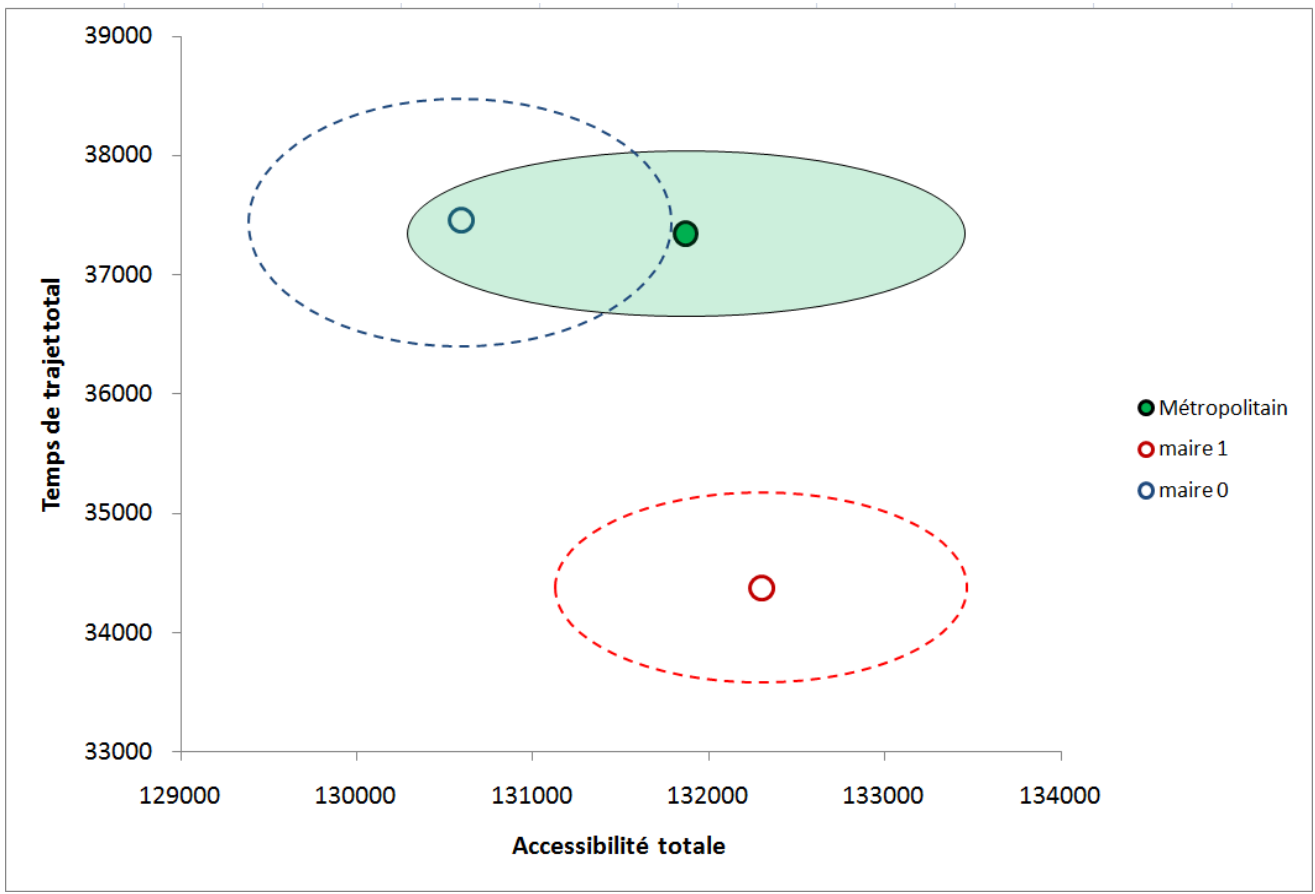

**Figure 4 : Outputs of the simulations : links between accessibility level and average travel times under different governance regimes.**

## Sensitivity analysis

Figure 5 shows the average accessibilities at the end of the simulation for four different initial configurations, making it possible to explore the most appropriate decision-making structures for a given urban form. Total accessibility is again taken as an output indicator; on the X-axis, the proportion of decisions taken locally shows, in the case of monocentric cities (the two cities are of unequal demographic importance), a clear tendency for centralized governance systems to offer better accessibility to the city. . Such a result is not reproduced in monocentric metropolises, where the two cities are of the same weight: the variable ξ seems independent of total accessibility even if it probably has an impact on the unequal spatial distribution of this accessibility.

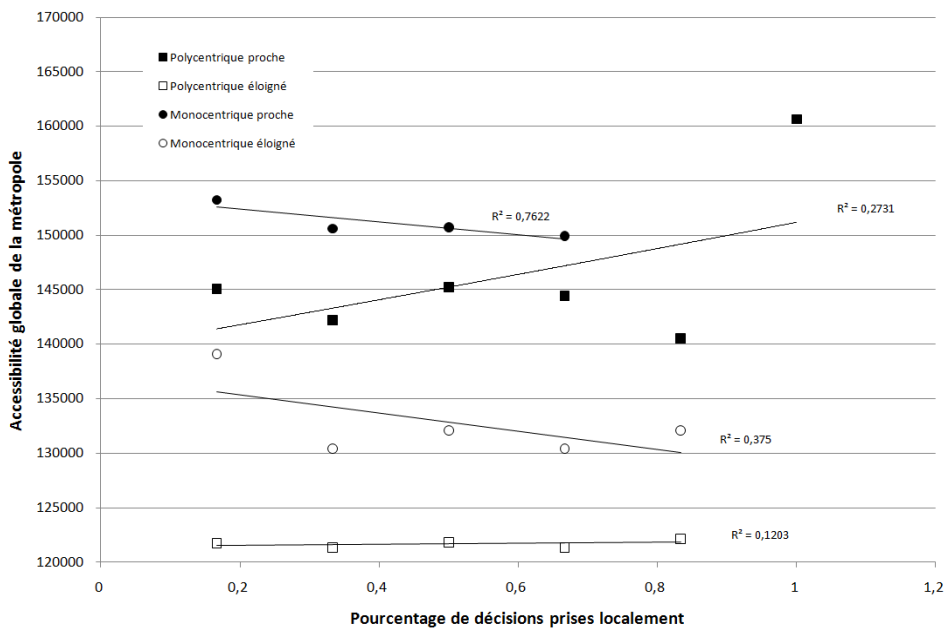

**Figure 5 : Sensitivity study: influence of urban form and type of governance on total accessibility level.**

# Discussion

The results presented constitute a first exploration of a model of interaction between transport and land use, proposing simultaneously to take into account the individual interests of location and displacement, and the collective interests of development. Of course, in its current simplistic configuration, and leave aside important aspects of the complexity of urban dynamics; the criteria with which transport infrastructures are built endogenously leave little room for the complex reality of the terrain. In addition, the evolution of land use has presented here only theoretically, because there is a challenge to adjust the temporalities of joint evolutions of the transport system and the settlement system.

So-called "complex systems" approaches aim at taking into account the succession of non-equilibrium states occurring in urban systems, for which neither individual nor global objective functions are available (Batty & Xie, 1999, Xie & Levinson, 2009, Cavailhes et al., 2010, Demare et al., 2017). Organized spatial configurations can emerge from a set of elementary rules, at the local level (as showed by Schelling's segregation model in the urban context (Gauvin et al., 2009). With such models, it is possible to obtain abrupt transitions as a result of the modification of one or a very small number of initial parameters. This is at the same time more realistic and more difficult to use in operational context for technical reasons such as the difficulty to calibrate such a model. Nevertheless we argue that such approaches are all the more necessary that we are in a period of multiple transitions that classical modelling tools have difficulty to capture.